\documentclass[letter]{aa}
\usepackage{ae,aecompl}

\usepackage{graphicx}

\usepackage{txfonts}
\usepackage{ulem}
\usepackage{hyperref}
\usepackage{graphicx}	
\usepackage{amsmath}	
\usepackage{amssymb}	
\usepackage{bm}
\usepackage{subfig}
\usepackage{mathtools}
\usepackage{float}
\usepackage{times}
\usepackage{color}
\usepackage{amsfonts}
\usepackage{booktabs}
\usepackage{siunitx}
\usepackage{soul}
\usepackage{xcolor}
\usepackage{amsmath}
\usepackage{float}
\usepackage{capt-of}
\usepackage{graphicx}
\usepackage{tikz}

\newcommand{\bhac}{\texttt{BHAC}~}

\begin{document}

\title{
Particle Acceleration via Transient Stagnation Surfaces in MADs During Flux Eruptions.}

\author{Vasilis Mpisketzis
          \inst{1,2}
          \and
          Georgios Filippos Paraschos
          \inst{3}
          \and
          Harry Ho-Yin Ng
          \inst{1}
          \and
          Antonios Nathanail
          \inst{4}
          }

\institute{Institut für Theoretische Physik, Goethe Universität Frankfurt, Max-von-Laue-Str.1, 60438 Frankfurt am Main, Germany
                \email{vmpisketzis@itp.uni-frankfurt.de}
            \and
            Department of Physics, National and Kapodistrian University of Athens, Panepistimiopolis, GR 15783 Zografos, Greece
            \and
            Max-Planck-Institut f\"ur Radioastronomie, Auf dem H\"ugel 69, D-53121 Bonn, Germany
            \and
            Research Center for Astronomy and Applied Mathematics, Academy of Athens, Soranou Efessiou 4, GR-11527 Athens, Greece
 }


\date{October}
\abstract
{In an accreting black hole, the appearance 
of a stagnation surface within the magnetized funnel 
has been analytically and numerically retrieved. 
However, in the complex physical environment during a flux eruption event, 
this situation may change drastically.
}
{
In this study, we focus on the simulation of accretion processes 
in Magnetically Arrested Disks (MADs) and investigate the dynamics
of plasma during flux eruption events.
}
{
We employ general relativistic magneto-hydrodynamic (GRMHD) simulations and 
search for regions with a divergent velocity during a flux eruption event. 
These regions would experience rapid and significant depletion of matter.
For this reason, we monitor the activation rate of the floor and the 
mass supply required for stable simulation evolution to further 
trace this transient stagnation surface. 
}
{
Our findings reveal an unexpected and persistent stagnation surface that develops 
during these eruptions, located around 2-3  gravitational radii 
(${\rm r_g}$) from the black hole. The stagnation surface is defined by a
divergent velocity field and is accompanied by enhanced mass addition.
This represents the first report of such a feature in this 
context. The stagnation surface is  ($7-9\,\,{\rm r_g}$) long. We estimate the overall potential difference along this stagnation surface for a 
supermassive black hole like M87 to be approximately 
$\Delta V \approx 10^{16}$ Volts.
}
{  
Our results indicate that, in MAD configurations, this transient stagnation surface during flux eruption events can be associated with  an accelerator of charged particles in the vicinity of supermassive black holes. In light of magnetic reconnection processes during these events, this work presents a complementary or an alternative mechanism for particle acceleration.
}

\keywords{relativstic processes, (stars:) gamma-ray burst: general, stars:neutron, black hole physics}

\maketitle

\section{Introduction} 

AGN-launched jets are powerful energy sources that lie in the center of galaxies. 
They are considered to originate from a black hole and are fueled by an accretion disk 
surrounding them \citep{2019ARA&A..57..467B}.
Magnetic fields play a key role in the jet formation, as jets are thought to be launched in connection to the accumulation of the magnetic flux that threads the event horizon \citep{M87_23, SgrA_24}. 
These field lines will rotate in the same direction as the black hole and provide the energy required for these jets as Poynting flux via the Blandford-Znajek mechanism \citep{1977MNRAS.179..433B}, as confirmed by GRMHD simulations \citep{2004MNRAS.350..427K,2011MNRAS.418L..79T} and observed in nature  in radio galaxies such as M\,87 \citep{Lu23} and 3C\,84 \citep{P23,P24b}. 

\begin{figure*}
\centering
\includegraphics[width=.83\textwidth]{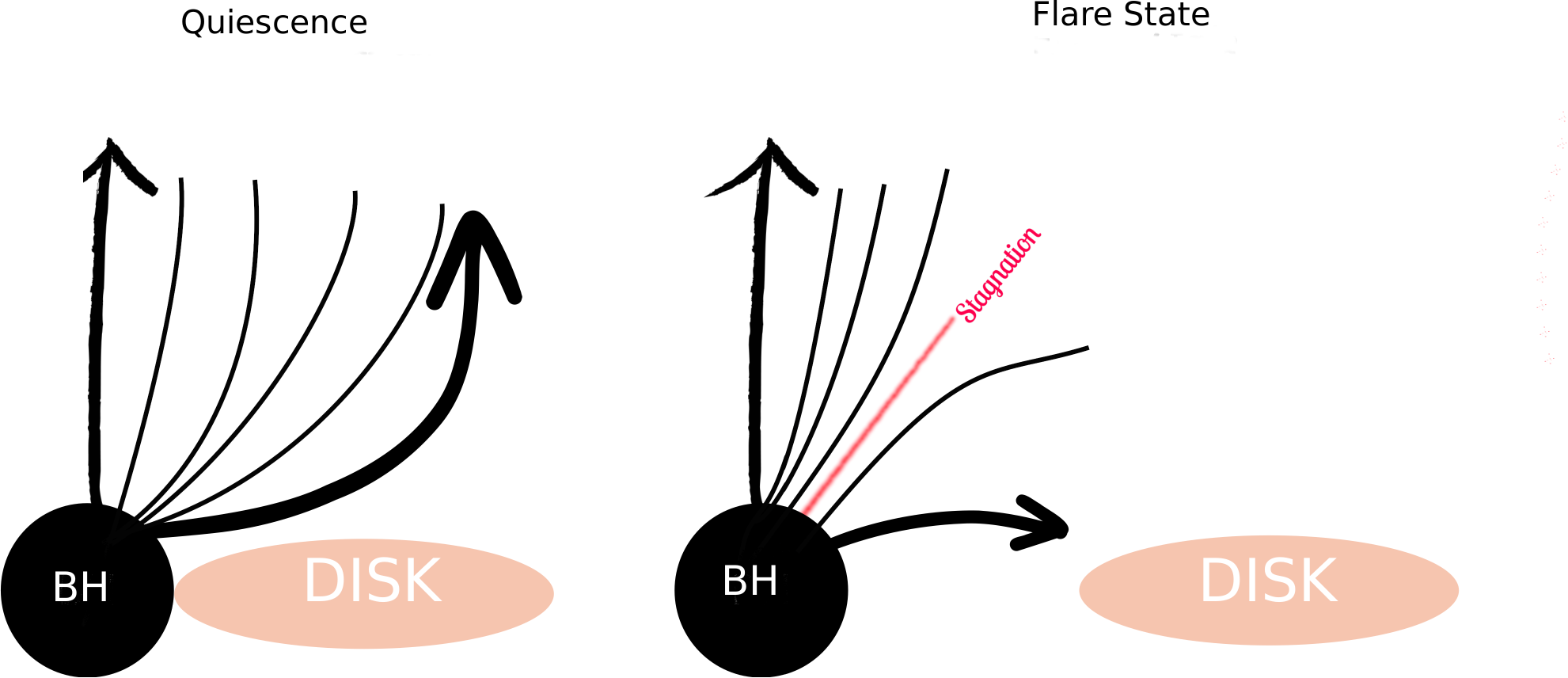}
\caption{Illustration of how a stagnation surface appears during flux eruption events. 
On the left panel, the funnel is displayed during normal evolution. On the right panel, the disk is expelled forcing the plasma in the boundary of the jet to follow a different trajectory  that can disconnect it from the inner funnel, ultimately resulting in the production of 
a transient stagnation surface (red).}
\label{cartoon}
\end{figure*}
However, unlike other compact objects \citep{1997ApJ...485..735M}, the black hole cannot provide the funnel with plasma. 
Quickly, all particles will either get accreted or ejected as an outflow, leaving it empty with no way of dynamically replenishing the plasma, since disk material field lines do not enter the jet region.
In this case, the ideal fluid approximation will break-down, introducing \(E_\parallel  \equiv \vec{E}\cdot \vec{B}\) which will accelerate non-thermal particles
to high energies \citep{2011ApJ...730..123L}. 
In nature, the mass replenishment is associated with physical processes that are able to create a substantial amount of electron-positron pairs within the funnel.  
For example, photons that are radiated from the accreting disk, are candidates for photon-photon collisions \citep{1996ApJ...468..330K,2010MNRAS.409.1183V,2011ApJ...735....9M}.  
Other mechanisms include particles that land within the funnel from the boundary of the jet. These are particle-particle interactions, or a combination of both particle-photon interactions.   
In a more complex case, a pair cascade  process (see for example \citealt{1977MNRAS.179..433B}) will be initiated and produce $\gamma$-ray emission.
This process takes place until the electric field is completely screened. This region can be recognized as a magnetospheric gap.  These gaps have been frequently associated with fast $\gamma$-ray variability of AGNs  \citep[see, for example,][]{2011ApJ...730..123L, 2014Sci...346.1080A, 2018ApJ...852..112K, 2019ApJ...883...66P, 2020ApJ...895..121C}.

In GRMHD simulations, certain regions are stabilized by manually 
resupplying each cell with the necessary plasma to maintain simulation 
stability \citep{2006MNRAS.368.1561M}. 
Recently, \cite{2024arXiv240401471C} introduced a combination of Force-Free 
and GRMHD solutions that are being smoothly connected from the wind 
region to the highly magnetized funnel, which however a priori assumes that the 
funnel region is filled with the required plasma.

For M87, a magnetically arrested disk \citep{2003PASJ...55L..69N} is considered
the most likely scenario \citep{2021ApJ...910L..13E}, where frequent magnetic reconnection events in the equatorial plane are believed to be responsible for high energy  radiation \citep{2023ApJ...943L..29H,2024arXiv240601211S}.

In this work we use GRMHD simulations to investigate the consistent and periodic appearance of a transient stagnation surface during the flux eruption events of a MAD system. 
They are defined as surfaces with a divergent velocity field. We further 
assist our efforts by monitoring the floor activation that takes 
place mainly inside the funnel.

The paper is organized as follows: in Sect.~\ref{sec:idea} we introduce the general idea of how we expect these stagnation surfaces to be formed, in Sect.~\ref{sec:setup}  we discuss the methods employed, in Sect.~\ref{sec:results}  we discuss the results and finally conclude in Sect.~\ref{sec:concl}.

\section{Physical concept of transient stagnation surface}\label{sec:idea}

It has been established \citep{2008ApJ...677..317I,2011MNRAS.418L..79T} that the MAD state is accompanied by large-amplitude fluctuations of magnetic flux and mass accretion. These fluctuations are caused by the quasi-periodic accumulation of excess magnetic flux that momentarily surpasses the ram pressure by the accretion disk. This results in blobs of magnetic flux that, starting from the vicinity of black hole, rotate and move outwards  due to the  tension of the magnetic fields \citep{2020MNRAS.497.4999D,2021MNRAS.502.2023P}.

During this process, the disk is briefly expelled in the azimuthal direction.
This has been studied extensively, revealing that the funnel takes a cylindrical shape and magnetic reconnection occurs \citep{2022ApJ...924L..32R}. At the onset of this event, the boundary of the funnel, near the equatorial plane, moves to larger cylindrical radii, while, near the axis of rotation, the velocity field is largely unaffected.  We suspect that such a physical environment can create a large velocity divergence in the event of strong flux eruptions.
The idea is schematically represented in  Fig. ~\ref{cartoon}.

It is expected that if this divergence occurs in the flow, a characteristic surface will appear that the motion of the local plasma will diverge, flowing in different directions.
Additionally,  the simulation will fail to preserve the maximum threshold of magnetization, due to the steep velocity gradient.  This surface is marked by the red line in Fig. ~\ref{cartoon}.
These main features differentiates this, from the inflow-outflow stagnation surface, where the transition is smooth in the radial velocity \citep{2006MNRAS.368.1561M} and its presence is continuous. 

In this investigation, we find it more effective to track and verify the occurrence of such 
physical configurations by monitoring the activation of the floor routine throughout the 
simulation, before examining the velocity field for the expected divergence.

\section{Numerical setup}
\label{sec:setup}
\begin{figure}

\centering
\includegraphics[width=.49\textwidth]{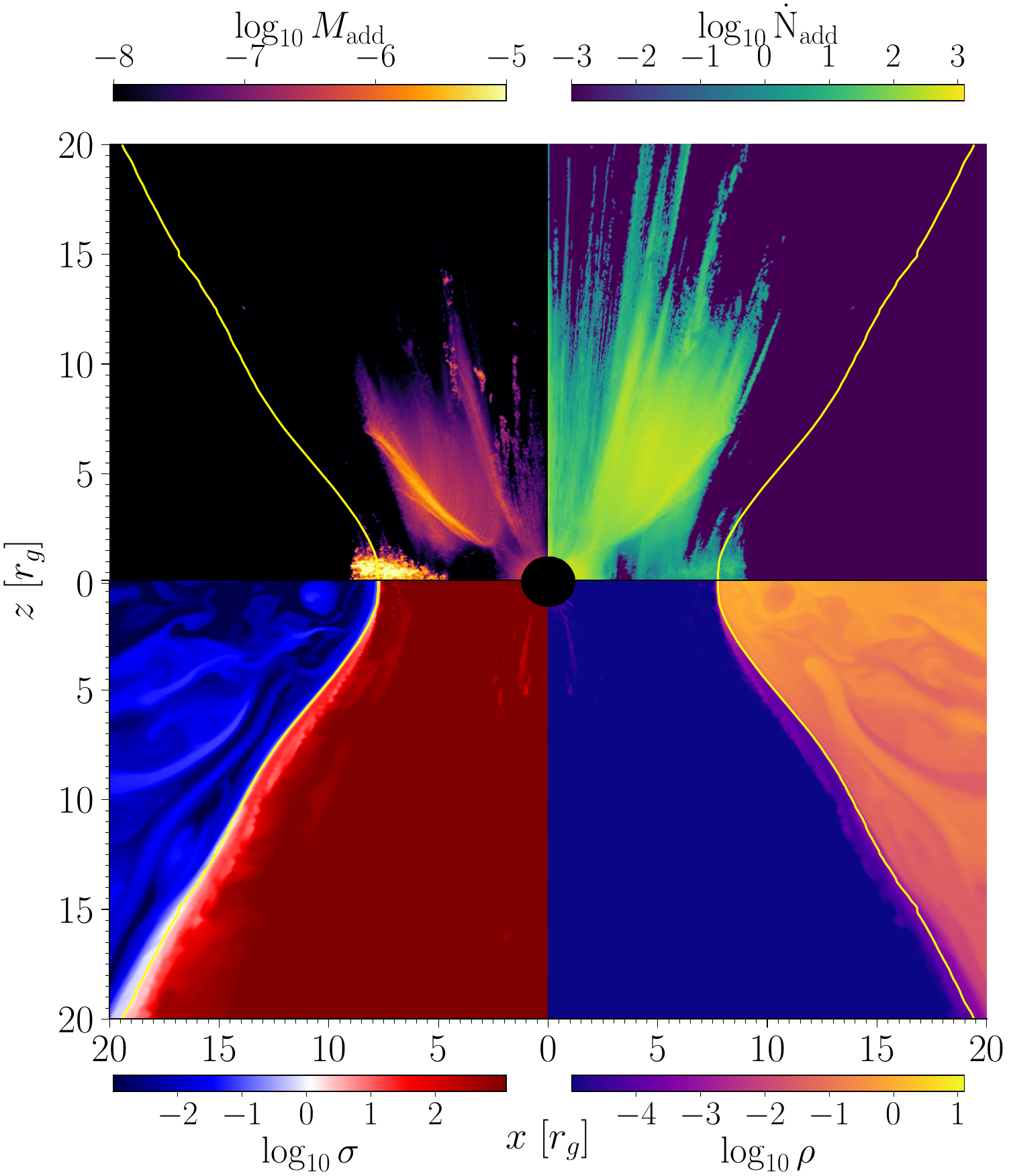}
\caption{ 
Top plot: Snapshot of the MAD simulation at $t=9200\mathrm{M}$. 
Top right and left panel:  $\dot{N}_\mathrm{add}$, and the the mass addition  by the floor, $M_\mathrm{add}$ respectively. 
Bottom right and left panel:  magnetization $\sigma$, and density $\rho$ respectively. 
}
\label{MAD0}
\end{figure}

\subsection{Initial conditions}
Throughout our work the simulations are carried out in two-dimensions ($\mathrm{2D}$), assuming azimuthal axisymmetry.
We performed our simulations employing the \bhac numerical code \citep{2017ComAC...4....1P, 2019A&A...629A..61O}, which solves the equations of the GRMHD framework using shock capturing methods.

Specifically, we initialized an equilibrium torus \citep{1976ApJ...207..962F}  with an inner radius of $r_{\mathrm{in}}=12$, and density maximum at $r_{\mathrm{max}}=25$ in Kerr black-hole spacetime. 
The dimensionless black hole spin is $\alpha = 0.9375$ 
and the boundaries in the $r-$coordinate were extended to $r_\mathrm{out} = 2500 r_g$.  
The initial vector potential is given by:
\begin{equation}
    A_{\phi} \propto \max \left[ \frac{\rho}{\rho{\text{max}}} \left( \frac{r}{r_{\text{in}}} \right)^3 \sin^3 \theta  \exp \left( -\frac{r}{400} \right) - 0.2, , 0 \right]
\end{equation}

which eventually leads to a MAD  \citep{2011MNRAS.418L..79T}. 
We use a resolution of $1536 \times 1024$ and evolve  until $t=10000 \ M$.

\subsection{Tracking} 
\label{sec:bhac}
In this subsection we provide details on the monitoring of the activation of 
the floor routine. This routine replenishes matter in regions that become more 
magnetized than the prescribed maximum magnetization for the simulation. 
Once the accretion starts, a highly magnetized funnel is built and 
additional mass is provided to the system to ensure the simulation stability.  
For each cell, if the magnetization $\sigma$ was higher than $\sigma_\mathrm{max}$, set to $\sigma_\mathrm{max} = 1000$ for our simulation, mass was added as described by the following equation:
\begin{equation}
    \rho_\mathrm{min} = \dfrac{b^2}{\sigma_\mathrm{max}},
\end{equation}
where $\sigma_\mathrm{max}$ is a threshold value set when initializing the simulation and $b^2$ is the square of the co-moving magnetic field of that cell;
\begin{equation}
    b^2 = \dfrac{B^2}{\Gamma ^2} + (\vec{B}\vec{v})^2.
\end{equation}


We calculated an extra variable $\mathcal{P}$, that cumulatively measures how much mass was generated due to the increased magnetization.
After each iteration this variable is updated according to the relation:
\begin{equation}
     \mathcal{P}_\mathrm{new} = \mathcal{P}_\mathrm{old} + \mathrm{dV}\big(\rho_\mathrm{min} - \rho_\mathrm{old}\big), \ \ \mathrm{if} \ \  \rho_\mathrm{old} < \rho_\mathrm{min},
\label{Padd}
\end{equation}

\begin{figure}

\centering
\includegraphics[width=.49\textwidth]{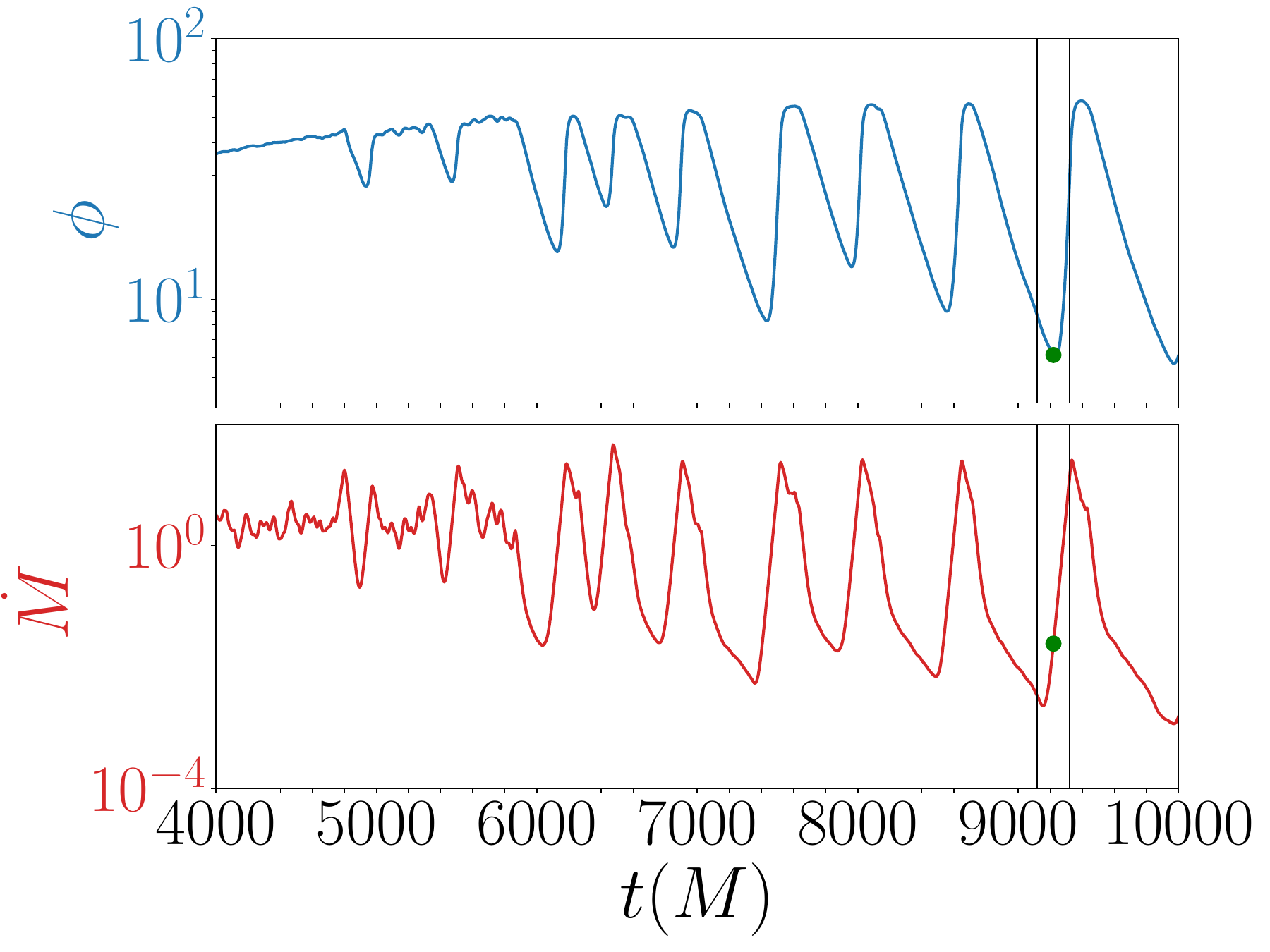}
\caption{ 
Evolution of the mass accretion rate (bottom) and magnetic flux (top)  integrated on the event horizon of the black hole.
The vertical black lines at  $t= 9100\mathrm{M},9300\mathrm{M}$, indicate the time interval in which we calculated the quantities of $\dot{N}_\mathrm{add}$ and $M_\mathrm{add}$ and the green dot represents the snapshot of the bottom panel of Fig. \ref{MAD0}.  
}
\label{MAD01}
\end{figure}
\begin{figure*}
\centering
\includegraphics[width=.7\textwidth]{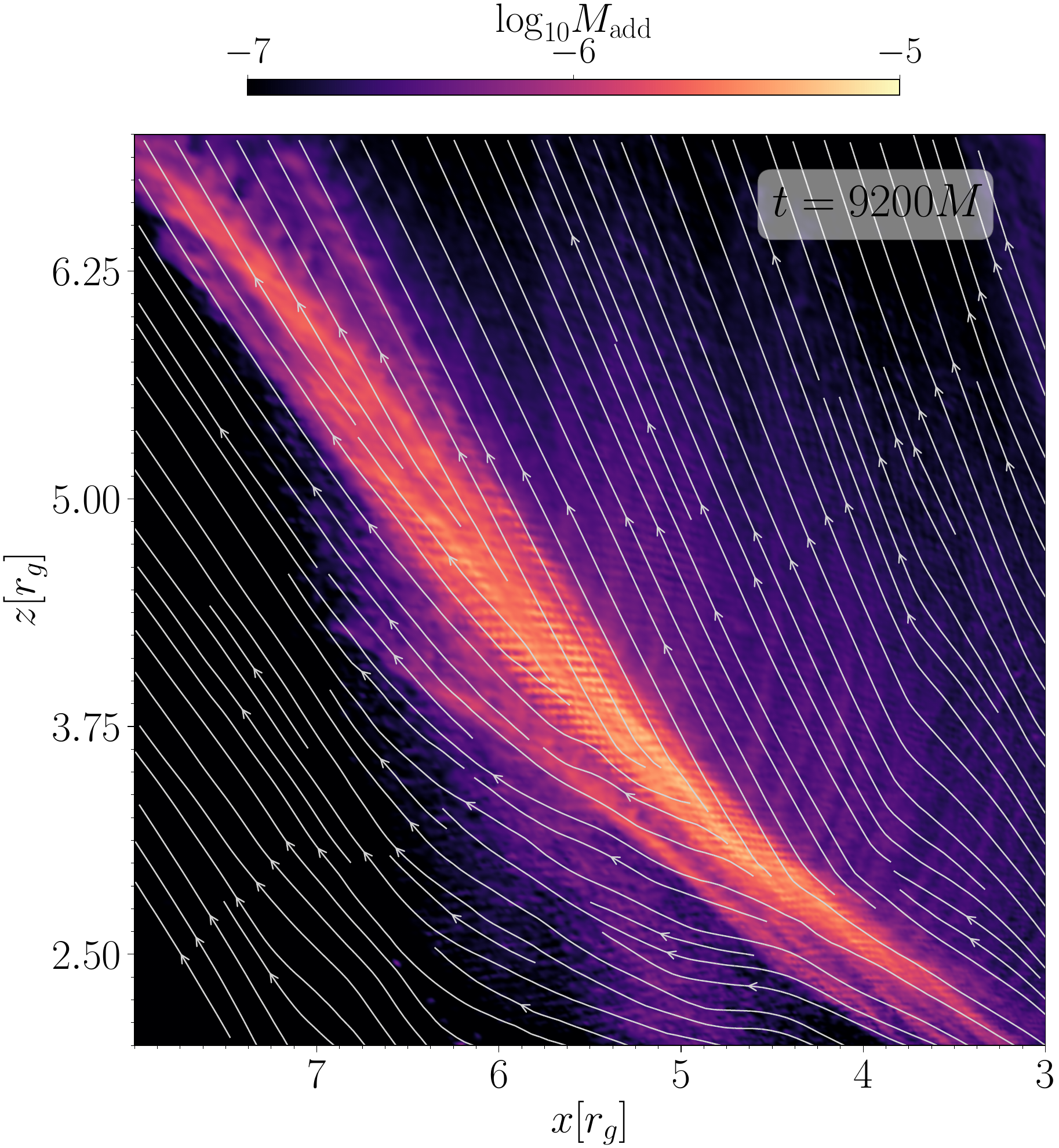}
\caption{Velocity field-lines of the stagnation surface at $t=9200 M$ in the transient stagnation surfaces during a flux eruption event. Background color depicts the mass addition by the floor routine. The velocity divergence is evident and coincides with the enhanced 
mass addition. }

\label{stag_vel}
\end{figure*}

\begin{figure*}
\centering
\includegraphics[width=.49\textwidth]{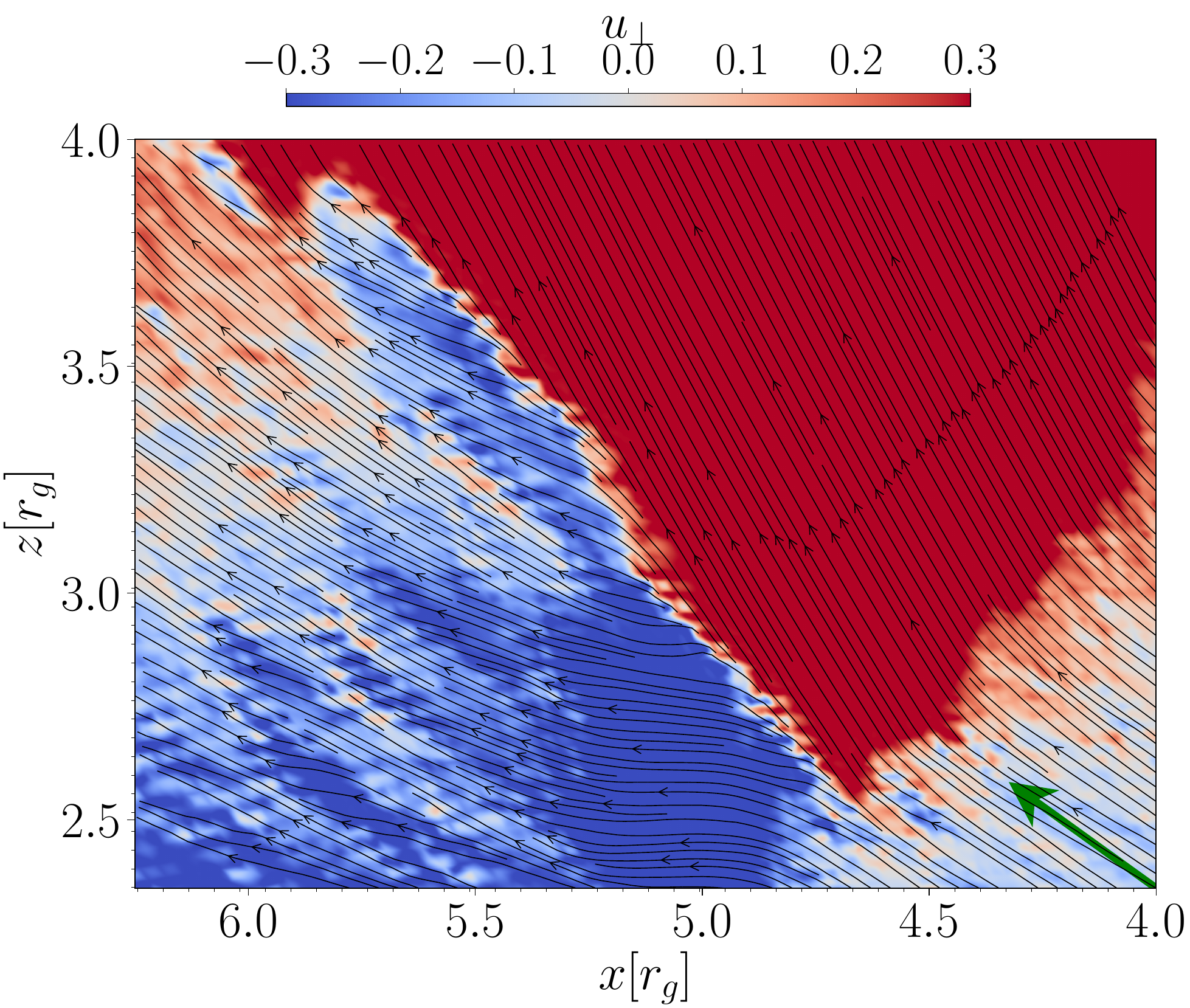}
\includegraphics[width=.49\textwidth]{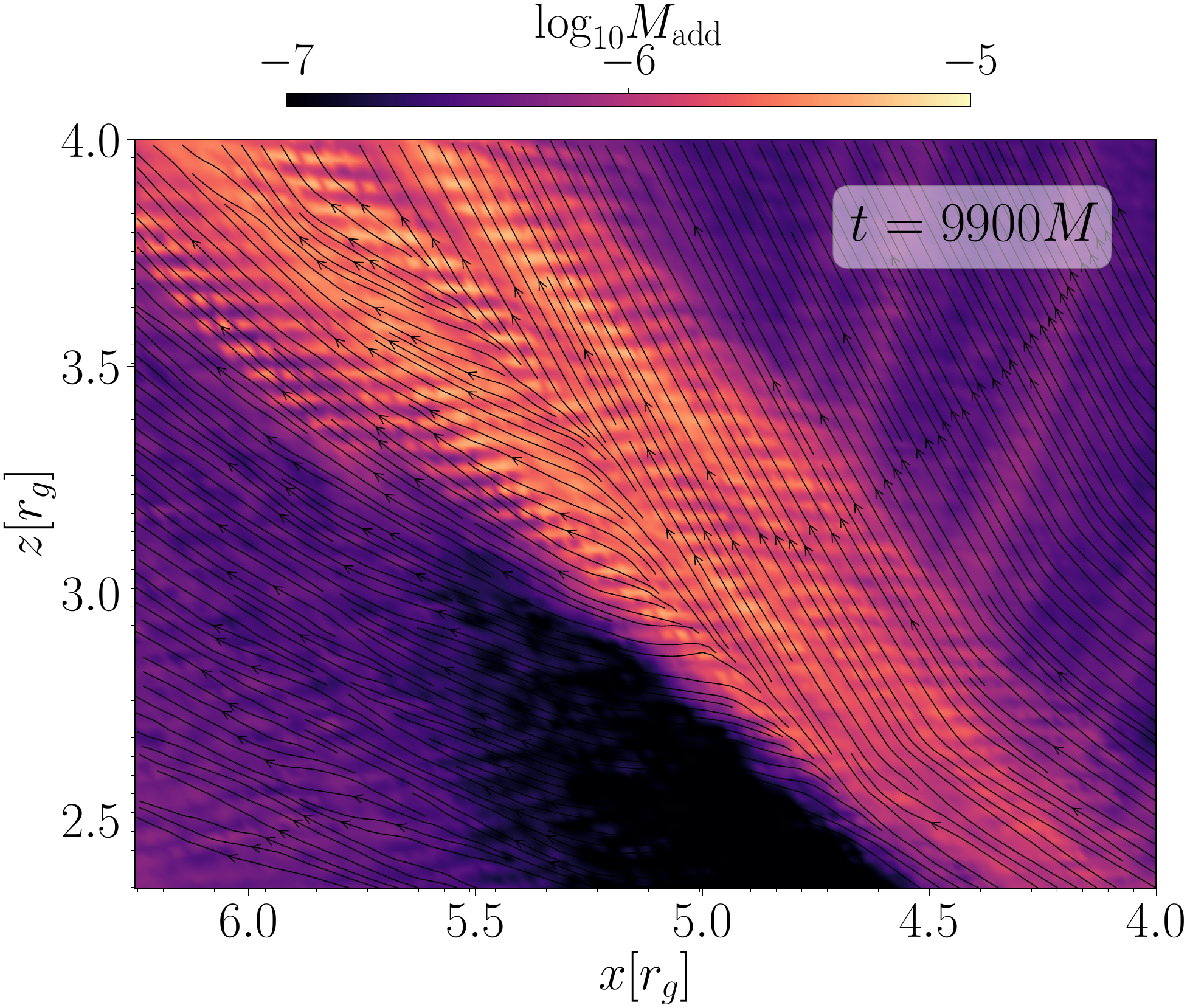}
\includegraphics[width=0.99\textwidth]{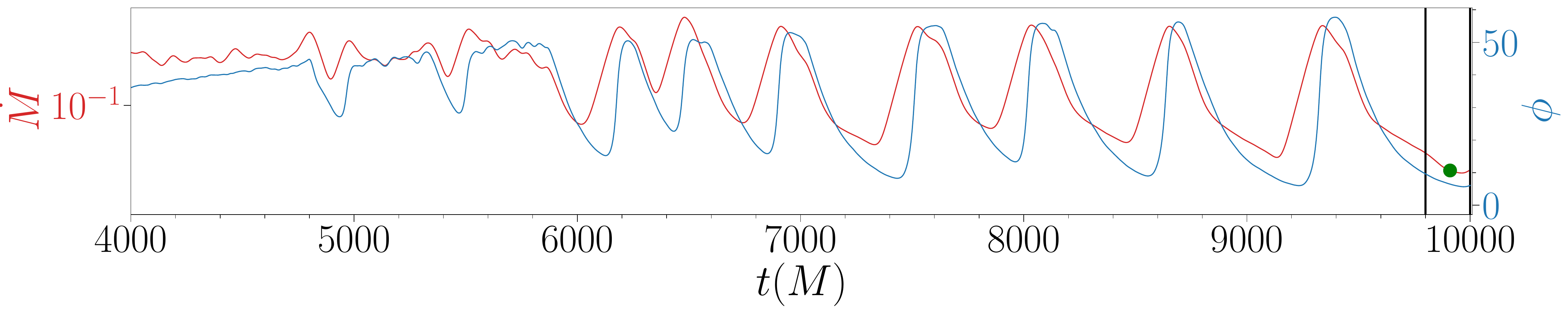 }
\caption{Velocity field-lines  at $t=9900 M$ in the transient stagnation surfaces during a
flux eruption event. Right panel:  the mass addition by the floor routine. Left panel:
the projection of the plasma velocity on the uniform flow vector (depicted on the bottom
corner in green). Bottom panel: Similar to Fig. ~\ref{MAD01}}

\label{stag_veltr}
\end{figure*}

where the indices $\mathrm{old}$ and $\mathrm{new}$ refer to each quantity's value before and after the floor corrections and $\mathrm{dV}$ to the volume of each cell.

Similarly, we measure the total number of floor activations using the relation: 
\begin{equation}
     \mathcal{N}_\mathrm{new} = \mathcal{N}_\mathrm{old} + 1,\ \ \mathrm{if} \  \rho_\mathrm{old} < \rho_\mathrm{min}.
\end{equation}

We initiated these variables at value $0$ and they cumulatively increase with simulation time. 
In order to measure a quantity that can provide more intuition, 
we selected a time span and calculated the mass addition:

\begin{equation}
    M_{\rm add} = \mathcal{P}_\mathrm{t_2} - \mathcal{P}_\mathrm{t_1}.
\end{equation}
Likewise, the floor activation rate is given by:

\begin{equation}
    \dot{N}_{\rm add} = \dfrac{\mathcal{N}_\mathrm{t_2} - \mathcal{N}_\mathrm{t_1}}{\mathrm{t_2}-\mathrm{t_1}},
\end{equation}
where $\mathrm{t_1}, \mathrm{t_2}$ are the start and stop times of the interval that we examined. 

Note that $\mathcal{P}, \mathcal{N}$ are monotonically increasing functions of time; therefore, both of these quantities are either greater or equal to zero. 

An implementation approach we explored was to measure these rates in the time-span of each iteration. 
This approach only yielded the cells that momentarily surpassed the magnetization threshold and did not provide further insight. 

\section{Results}\label{sec:results}
\subsection{Flux eruption events}
During the time-span of $10000 \mathrm{M}$  there have been multiple instances of flux eruptions as hinted by the magnetic flux accumulated on the black hole and the rest-mass accretion rate  shown in Fig. ~\ref{MAD01}. Due to the nature of 2D simulations, the whole accretion process ceases during these events, and only in 3D 
simulation accretion can continue through non-axisymmetric instabilities
\citep{Papadopoulos2019}.

In the upper panel of Fig. ~\ref{MAD0} the following quantities are presented, 
magnetization $\sigma$,  density $\rho$, $\dot{N}_\mathrm{add}$, and $M_\mathrm{add}$ during a standard magnetic-flux eruption event at $t=9200\mathrm{M}$. During the event
the disk gets momentarily expelled. As is evident by the magnetization and density, in the bottom panels, no fluctuations are visible within the funnel. This however, changes drastically, when examining the introduced quantities  in the previous section.
In the left panel (showing the mass addition), visible structures have been formed during the whole time span of the flux eruption event, seen within the funnel and near the equatorial plane, where the boundary has moved significantly. 
We rely on these structures to further examine if such locations constitute  the transient  stagnation surfaces.


\subsection{Transient stagnation surface}

To do so, in Fig. ~\ref{stag_vel} we zoom into such a region and plot the velocity streamlines with increased density. This reveals the divergence in the velocity that completely coincides with the  surface of enhanced mass addition. 
In a similar event in particle-in-cell accretion simulations
\citep{2023PhRvL.130k5201G}, this kind of velocity field was not retrieved, or at least was not 
highlighted. This was most probably due to  the existence of critical differences, such as  the 
substitution of the accretion disk with a Bondi-accretion and the shorter simulation time in 
that work. 

The investigation of other flux eruption events indicate the periodic nature of the 
formation of transient stagnation surfaces that accompany each event. 
To this end, in Fig. ~\ref{stag_veltr} we showcase a different flux eruption event, at $t=9900\mathrm{M}$. In the right plot, we display the
projection of the velocity on the uniform flow $u_\perp = \hat{u} \cdot \hat{v}_{\perp}$, prior to the divergence, indicated by the green arrow in the bottom right panel,
where $\hat{v}_\perp$ is the unit vector $(-0.6,0.8)$, perpendicular to the flow that precedes the transient stagnation surface. For that reason, the values prior to the divergence, are fluctuations of small values around zero. A strong discontinuity in this velocity component is present, centered at the stagnation surface. It coincides with a region of enhanced mass addition, as in the previous eruption event. 
The reported transient stagnation surfaces have a length of $7-9\,\, {\rm r_g}$ and are well fitted by;
\begin{equation}
    y_\mathrm{stag,9200} = 0.98 x_\mathrm{stag,9200} - 1.5
\end{equation}
and
\begin{equation}
    y_\mathrm{stag,9900} = 1.08 x_\mathrm{stag,9900} - 2.5
\end{equation}
verifying their radial geometry.

During flux eruption events, the depletion of plasma at stagnation surfaces can 
lead to the rapid development of a strong electric field parallel to the magnetic 
field, capable of accelerating particles to ultra-relativistic energies. For M87, 
assuming that the electric field reaches values close to the magnetic field 
strength (\( B \approx 5 - 100 \, \mathrm{G} \); \citep{EHT_M87_PaperV}), the 
resulting electric field is \( E_{\mathrm{stag}} \approx 10 \, \mathrm{statvolt/
cm} = 3 \times 10^5 \, \mathrm{V/m} \). This produces an estimated potential 
difference:
\begin{equation}
\Delta V \approx 3 \times 10^{16} \left( \frac{E}{3 \times 10^6 \, \mathrm{V/m}} \right) \left( \frac{h}{0.1 \, r_{g, \, \mathrm{M87}}} \right) \, \mathrm{V},
\end{equation}
where $h$ is the magnetospheric gap length \citep{2016A&A...593A...8P}. The power per particle accelerated through this potential difference \(\Delta V\) 
is given by the product of the particle charge and the voltage, which yields 
\(4.8 \, \text{MeV}\) per particle.
For a particle injection rate of, say, \(\dot{N} \approx 10^{42} \, \text{s}^{-1}\), the total accelerating power is:
\[
P_{\text{stag}}\approx 4.8 \times 10^{46} \, \text{erg} \, \text{s}^{-1}.
\]
To have a feeling of the particle injection rate, 
we estimate the Goldreich-Julian number near the M87 black hole 
to be \(N_{\text{GJ}} \approx 5.5 \times 10^{49}\)\footnote{The  
Goldreich-Julian number density is \( n_{\text{GJ}} = 
\frac{\Omega B}{2 \pi e c} \), where \( \Omega \approx \frac{a c}{2 r_g}\) is the 
angular frequency of the black hole (for a rapidly rotating black hole
the dimensionless spin $a\approx 1$) and \( e \) 
the elementary charge. Near M87, \(n_{\text{GJ}} \approx 8.3 \times 10^5 \, 
\text{cm}^{-3}\), which at the stagnation surface for a typical length of 
$7-9\,\, {\rm r_g}$, and a width of $0.1\,\, {\rm r_g}$ 
( a volume of \(V \approx 0.7 \, r_g^3\) ), yields an amount of 
\(N_{\text{GJ}} \approx 5.5 \times 10^{49}\) particles. }. 
Thus, a tiny fraction of 
these particles (e.g., $<0.0001\%$)  would contribute a powerful injection rate.

Matching charge acceleration with radiation losses, the maximum Lorentz factor \( 
\gamma \) of particles is set by synchrotron and inverse Compton (IC) cooling 
limits. Synchrotron losses give:
\begin{equation}
\gamma_{\text{synch}} = 3 \times 10^7 B_1^{-0.5},
\end{equation}
where \( B_1 = B / (10 \, \mathrm{G}) \), while IC losses in 
the intense photon field of the inner jet constrain \( \gamma \) to
\begin{equation}
\gamma_{\text{IC}} = 2 \times 10^6 B_1^{1/2} M_9 L_{43}^{-1/2}
\mathcal{R} (h / r_s)^{1/2},
\end{equation}
where \( M_9 = M / (10^9 M_\odot) \), \( \mathcal{R} = R_d / r_g \) is the 
rescaled size of the emission region \( R_d \), and \( L_{43} = L_{\mathrm{d}} / 
10^{43} \, \mathrm{erg \, s^{-1}} \) is the disk’s photon luminosity 
\citep{2011ApJ...730..123L}. While the curvature radiation limit on \( \gamma \) would be less 
restrictive in this environment, synchrotron and IC losses are dominant for 
particles accelerated in the stagnation surface, effectively setting the maximum 
energies.

Given the strong magnetic field and high acceleration rates, synchrotron emission 
is expected to peak at the MeV scale, while inverse Compton scattering could 
produce TeV-range photons. These conditions suggest that transient TeV flares in 
M87’s jet are plausible during flux eruption events, as accelerated particles 
rapidly cool via IC and synchrotron radiation under the intense acceleration
in a stagnation surface\footnote{Note that, TeV photons face significant opacity 
constraints; photon-photon collisions with background photons (e.g., in the 1 eV 
range) near the horizon can produce electron-positron pairs, likely impeding the 
escape of TeV photons under typical conditions \citep{2011ApJ...730..123L}.}.

 \section{Conclusions}\label{sec:concl}
In this work, we examined a standard MAD scenario with the aim of detecting stagnation surfaces 
during magnetic flux eruptions, defined as regions with divergent velocity. To achieve this, we 
introduced a new diagnostic quantity that tracks the activation rate of the floor routine, 
which injects mass into the system whenever magnetization exceeds a certain threshold. Within 
the highly magnetized funnel, this enabled us to trace the formation of stagnation surfaces 
during flux eruptions. We highlight two such events, showing dense velocity streamlines and a 
discontinuity in the velocity component perpendicular to the preceding flow. This discontinuity 
aligns with enhanced mass injection by the floor routine, further validating our tracking
approach. Unlike the typical inflow-outflow stagnation surface, The scenario we observed reveals a near-radial stagnation surface, forming only during flux eruptions due to the outwards move of the funnel in the equatorial plane. Such surfaces could potentially be responsible for high-energy particle acceleration \citep{2011ApJ...730..123L}.

A more accurate representation of this phenomenon would require 3D simulations, where tracking 
the stagnation surface becomes significantly more challenging due to the increased resolution 
needed and the difficulty of tracking it. It would necessitate a curved slice aligned with 
the low-magnetized regions formed by flux eruptions \citep{2022ApJ...941...30C}. The mass 
addition  tracking introduced in this study is well-suited for 3D simulations as well. 
In our current simulations, we were limited by relatively small $\sigma_\mathrm{max}$ values, 
constraints that become even more stringent in 3D. However, recent developments in numerical 
codes \citep{2024PhRvD.109f4061N} offer greater tolerance in handling such limitations, which 
could improve the detection and validation of this concept.

\section*{Acknowledgements}
We thank L. Rezzolla, S. I. Stathopoulos, A. Loules, A. Tursunov and A. Cruz-Osorio   for useful discussions.
Support also comes from the ERC Advanced Grant
``JETSET: Launching, propagation and emission of relativistic jets from
binary mergers and across mass scales'' (Grant No. 884631),
and the ERC advanced grant “M2FINDERS - Mapping Magnetic Fields with INterferometry Down to Event hoRizon Scales” (Grant No. 101018682).
\section*{Data Availability}
The data underlying this article will be shared on reasonable request to the corresponding author.


\bibliographystyle{aa}
\bibliography{other2}

\begin{thebibliography}{37}
\expandafter\ifx\csname natexlab\endcsname\relax\def\natexlab#1{#1}\fi

\bibitem[{{Aleksi{\'c}} {et~al.}(2014){Aleksi{\'c}}, {Ansoldi}, {Antonelli},
  {Antoranz}, {Babic}, {Bangale}, {Barrio}, {Gonz{\'a}lez}, {Bednarek},
  {Bernardini}, {Biasuzzi}, {Biland}, {Blanch}, {Bonnefoy}, {Bonnoli},
  {Borracci}, {Bretz}, {Carmona}, {Carosi}, {Colin}, {Colombo}, {Contreras},
  {Cortina}, {Covino}, {Da Vela}, {Dazzi}, {De Angelis}, {De Caneva}, {De
  Lotto}, {Wilhelmi}, {Mendez}, {Prester}, {Dorner}, {Doro}, {Einecke},
  {Eisenacher}, {Elsaesser}, {Fonseca}, {Font}, {Frantzen}, {Fruck}, {Galindo},
  {L{\'o}pez}, {Garczarczyk}, {Terrats}, {Gaug}, {Godinovi{\'c}}, {Mu{\~n}oz},
  {Gozzini}, {Hadasch}, {Hanabata}, {Hayashida}, {Herrera}, {Hildebrand},
  {Hose}, {Hrupec}, {Idec}, {Kadenius}, {Kellermann}, {Kodani}, {Konno},
  {Krause}, {Kubo}, {Kushida}, {La Barbera}, {Lelas}, {Lewandowska},
  {Lindfors}, {Lombardi}, {Longo}, {L{\'o}pez}, {L{\'o}pez-Coto},
  {L{\'o}pez-Oramas}, {Lorenz}, {Lozano}, {Makariev}, {Mallot}, {Maneva},
  {Mankuzhiyil}, {Mannheim}, {Maraschi}, {Marcote}, {Mariotti},
  {Mart{\'\i}nez}, {Mazin}, {Menzel}, {Miranda}, {Mirzoyan}, {Moralejo},
  {Munar-Adrover}, {Nakajima}, {Niedzwiecki}, {Nilsson}, {Nishijima}, {Noda},
  {Orito}, {Overkemping}, {Paiano}, {Palatiello}, {Paneque}, {Paoletti},
  {Paredes}, {Paredes-Fortuny}, {Persic}, {Poutanen}, {Moroni}, {Prandini},
  {Puljak}, {Reinthal}, {Rhode}, {Rib{\'o}}, {Rico}, {Garcia}, {R{\"u}gamer},
  {Saito}, {Saito}, {Satalecka}, {Scalzotto}, {Scapin}, {Schultz}, {Schweizer},
  {Shore}, {Sillanp{\"a}{\"a}}, {Sitarek}, {Snidaric}, {Sobczynska}, {Spanier},
  {Stamatescu}, {Stamerra}, {Steinbring}, {Storz}, {Strzys}, {Takalo},
  {Takami}, {Tavecchio}, {Temnikov}, {Terzi{\'c}}, {Tescaro}, {Teshima},
  {Thaele}, {Tibolla}, {Torres}, {Toyama}, {Treves}, {Uellenbeck}, {Vogler},
  {Zanin}, {Kadler}, {Schulz}, {Ros}, {Bach}, {Krau{\ss}}, \&
  {Wilms}}]{2014Sci...346.1080A}
{Aleksi{\'c}}, J., {Ansoldi}, S., {Antonelli}, L.~A., {et~al.} 2014, Science,
  346, 1080

\bibitem[{{Blandford} {et~al.}(2019){Blandford}, {Meier}, \&
  {Readhead}}]{2019ARA&A..57..467B}
{Blandford}, R., {Meier}, D., \& {Readhead}, A. 2019, \araa, 57, 467

\bibitem[{{Blandford} \& {Znajek}(1977)}]{1977MNRAS.179..433B}
{Blandford}, R.~D. \& {Znajek}, R.~L. 1977, \mnras, 179, 433

\bibitem[{{Chael}(2024)}]{2024arXiv240401471C}
{Chael}, A. 2024, arXiv e-prints, arXiv:2404.01471

\bibitem[{{Chatterjee} \& {Narayan}(2022)}]{2022ApJ...941...30C}
{Chatterjee}, K. \& {Narayan}, R. 2022, \apj, 941, 30

\bibitem[{{Chen} \& {Yuan}(2020)}]{2020ApJ...895..121C}
{Chen}, A.~Y. \& {Yuan}, Y. 2020, \apj, 895, 121

\bibitem[{{Dexter} {et~al.}(2020){Dexter}, {Tchekhovskoy},
  {Jim{\'e}nez-Rosales}, {Ressler}, {Baub{\"o}ck}, {Dallilar}, {de Zeeuw},
  {Eisenhauer}, {von Fellenberg}, {Gao}, {Genzel}, {Gillessen}, {Habibi},
  {Ott}, {Stadler}, {Straub}, \& {Widmann}}]{2020MNRAS.497.4999D}
{Dexter}, J., {Tchekhovskoy}, A., {Jim{\'e}nez-Rosales}, A., {et~al.} 2020,
  \mnras, 497, 4999

\bibitem[{{Event Horizon Telescope Collaboration} {et~al.}(2024){Event Horizon
  Telescope Collaboration}, {Akiyama}, {Alberdi}, {Alef}, {Algaba}, {Anantua},
  {Asada}, {Azulay}, {Bach}, {Baczko}, {Ball}, {Balokovic}, {Bandyopadhyay},
  {Barrett}, {Baub{\"o}ck}, {Benson}, {Bintley}, {Blackburn}, {Blundell},
  {Bouman}, {Bower}, {Boyce}, {Bremer}, {Brinkerink}, {Brissenden}, {Britzen},
  {Broderick}, {Broguiere}, {Bronzwaer}, {Bustamante}, {Byun}, {Carlstrom},
  {Ceccobello}, {Chael}, {Chan}, {Chang}, {Chatterjee}, {Chatterjee}, {Chen},
  {Chen}, {Cheng}, {Cho}, {Christian}, {Conroy}, {Conway}, {Cordes},
  {Crawford}, {Crew}, {Cruz-Osorio}, {Cui}, {Dahale}, {Davelaar}, {De
  Laurentis}, {Deane}, {Dempsey}, {Desvignes}, {Dexter}, {Dhruv}, {Dihingia},
  {Doeleman}, {Dougal}, {Dzib}, {Eatough}, {Emami}, {Falcke}, {Farah}, {Fish},
  {Fomalont}, {Ford}, {Foschi}, {Fraga-Encinas}, {Freeman}, {Friberg}, {Fromm},
  {Fuentes}, {Galison}, {Gammie}, {Garc{\'\i}a}, {Gentaz}, {Georgiev}, {Goddi},
  {Gold}, {G{\'o}mez-Ruiz}, {G{\'o}mez}, {Gu}, {Gurwell}, {Hada}, {Haggard},
  {Haworth}, {Hecht}, {Hesper}, {Heumann}, {Ho}, {Ho}, {Honma}, {Huang},
  {Huang}, {Hughes}, {Ikeda}, {Impellizzeri}, {Inoue}, {Issaoun}, {James},
  {Jannuzi}, {Janssen}, {Jeter}, {Jiang}, {Jim{\'e}nez-Rosales}, {Johnson},
  {Jorstad}, {Joshi}, {Jung}, {Karami}, {Karuppusamy}, {Kawashima}, {Keating},
  {Kettenis}, {Kim}, {Kim}, {Kim}, {Kim}, {Kino}, {Koay}, {Kocherlakota},
  {Kofuji}, {Koch}, {Koyama}, {Kramer}, {Kramer}, {Kramer}, {Krichbaum}, {Kuo},
  {La Bella}, {Lauer}, {Lee}, {Lee}, {Leung}, {Levis}, {Li}, {Lico}, {Lindahl},
  {Lindqvist}, {Lisakov}, {Liu}, {Liu}, {Liuzzo}, {Lo}, {Lobanov}, {Loinard},
  {Lonsdale}, {Lowitz}, {Lu}, {MacDonald}, {Mao}, {Marchili}, {Markoff},
  {Marrone}, {Marscher}, {Mart{\'\i}-Vidal}, {Matsushita}, {Matthews},
  {Medeiros}, {Menten}, {Michalik}, {Mizuno}, {Mizuno}, {Moran}, {Moriyama},
  {Moscibrodzka}, {Mulaudzi}, {M{\"u}ller}, {M{\"u}ller}, {Mus}, {Musoke},
  {Myserlis}, {Nadolski}, {Nagai}, {Nagar}, {Nakamura}, {Narayanan},
  {Natarajan}, {Nathanail}, {Fuentes}, {Neilsen}, {Neri}, {Ni}, {Noutsos},
  {Nowak}, {Oh}, {Okino}, {Olivares}, {Ortiz-Le{\'o}n}, {Oyama}, {{\"O}zel},
  {Palumbo}, {Paraschos}, {Park}, {Parsons}, {Patel}, {Pen}, {Pesce},
  {Pi{\'e}tu}, {Plambeck}, {PopStefanija}, {Porth}, {P{\"o}tzl}, {Prather},
  {Preciado-L{\'o}pez}, {Psaltis}, {Pu}, {Ramakrishnan}, {Rao}, {Rawlings},
  {Raymond}, {Rezzolla}, {Ricarte}, {Ripperda}, {Roelofs}, {Rogers},
  {Romero-Ca{\~n}izales}, {Ros}, {Roshanineshat}, {Rottmann}, {Roy}, {Ruiz},
  {Ruszczyk}, {Rygl}, {S{\'a}nchez}, {S{\'a}nchez-Arg{\"u}elles},
  {S{\'a}nchez-Portal}, {Sasada}, {Satapathy}, {Savolainen}, {Schloerb},
  {Schonfeld}, {Schuster}, {Shao}, {Shen}, {Small}, {Sohn}, {SooHoo},
  {Sosapanta Salas}, {Souccar}, {Stanway}, {Sun}, {Tazaki}, {Tetarenko},
  {Tiede}, {Tilanus}, {Titus}, {Torne}, {Toscano}, {Traianou}, {Trent},
  {Trippe}, {Turk}, {van Bemmel}, {van Langevelde}, {van Rossum}, {Vos},
  {Wagner}, {Ward-Thompson}, {Wardle}, {Washington}, {Weintroub}, {Wharton},
  {Wielgus}, {Wiik}, {Witzel}, {Wondrak}, {Wong}, {Wu}, {Yadlapalli},
  {Yamaguchi}, {Yfantis}, {Yoon}, {Young}, {Young}, {Younsi}, {Yu}, {Yuan},
  {Yuan}, {Zensus}, {Zhang}, {Zhao}, \& {Zhao}}]{SgrA_24}
{Event Horizon Telescope Collaboration}, {Akiyama}, K., {Alberdi}, A., {et~al.}
  2024, \apjl, 964, L25

\bibitem[{{Event Horizon Telescope Collaboration} {et~al.}(2023){Event Horizon
  Telescope Collaboration}, {Akiyama}, {Alberdi}, {Alef}, {Algaba}, {Anantua},
  {Asada}, {Azulay}, {Bach}, {Baczko}, {Ball}, {Balokovi{\'c}}, {Barrett},
  {Baub{\"o}ck}, {Benson}, {Bintley}, {Blackburn}, {Blundell}, {Bouman},
  {Bower}, {Boyce}, {Bremer}, {Brinkerink}, {Brissenden}, {Britzen},
  {Broderick}, {Broguiere}, {Bronzwaer}, {Bustamante}, {Byun}, {Carlstrom},
  {Ceccobello}, {Chael}, {Chan}, {Chang}, {Chatterjee}, {Chatterjee}, {Chen},
  {Chen}, {Cheng}, {Cho}, {Christian}, {Conroy}, {Conway}, {Cordes},
  {Crawford}, {Crew}, {Cruz-Osorio}, {Cui}, {Dahale}, {Davelaar}, {De
  Laurentis}, {Deane}, {Dempsey}, {Desvignes}, {Dexter}, {Dhruv}, {Doeleman},
  {Dougal}, {Dzib}, {Eatough}, {Emami}, {Falcke}, {Farah}, {Fish}, {Fomalont},
  {Ford}, {Foschi}, {Fraga-Encinas}, {Freeman}, {Friberg}, {Fromm}, {Fuentes},
  {Galison}, {Gammie}, {Garc{\'\i}a}, {Gentaz}, {Georgiev}, {Goddi}, {Gold},
  {G{\'o}mez-Ruiz}, {G{\'o}mez}, {Gu}, {Gurwell}, {Hada}, {Haggard}, {Haworth},
  {Hecht}, {Hesper}, {Heumann}, {Ho}, {Ho}, {Honma}, {Huang}, {Huang},
  {Hughes}, {Ikeda}, {Impellizzeri}, {Inoue}, {Issaoun}, {James}, {Jannuzi},
  {Janssen}, {Jeter}, {Jiang}, {Jim{\'e}nez-Rosales}, {Johnson}, {Jorstad},
  {Joshi}, {Jung}, {Karami}, {Karuppusamy}, {Kawashima}, {Keating}, {Kettenis},
  {Kim}, {Kim}, {Kim}, {Kim}, {Kino}, {Koay}, {Kocherlakota}, {Kofuji}, {Koch},
  {Koyama}, {Kramer}, {Kramer}, {Kramer}, {Krichbaum}, {Kuo}, {La Bella},
  {Lauer}, {Lee}, {Lee}, {Leung}, {Levis}, {Li}, {Lico}, {Lindahl},
  {Lindqvist}, {Lisakov}, {Liu}, {Liu}, {Liuzzo}, {Lo}, {Lobanov}, {Loinard},
  {Lonsdale}, {Lowitz}, {Lu}, {MacDonald}, {Mao}, {Marchili}, {Markoff},
  {Marrone}, {Marscher}, {Mart{\'\i}-Vidal}, {Matsushita}, {Matthews},
  {Medeiros}, {Menten}, {Michalik}, {Mizuno}, {Mizuno}, {Moran}, {Moriyama},
  {Moscibrodzka}, {Mulaudzi}, {M{\"u}ller}, {M{\"u}ller}, {Mus}, {Musoke},
  {Myserlis}, {Nadolski}, {Nagai}, {Nagar}, {Nakamura}, {Narayan}, {Narayanan},
  {Natarajan}, {Nathanail}, {Fuentes}, {Neilsen}, {Neri}, {Ni}, {Noutsos},
  {Nowak}, {Oh}, {Okino}, {Olivares}, {Ortiz-Le{\'o}n}, {Oyama}, {{\"O}zel},
  {Palumbo}, {Paraschos}, {Park}, {Parsons}, {Patel}, {Pen}, {Pesce},
  {Pi{\'e}tu}, {Plambeck}, {PopStefanija}, {Porth}, {P{\"o}tzl}, {Prather},
  {Preciado-L{\'o}pez}, {Psaltis}, {Pu}, {Ramakrishnan}, {Rao}, {Rawlings},
  {Raymond}, {Rezzolla}, {Ricarte}, {Ripperda}, {Roelofs}, {Rogers},
  {Romero-Ca{\~n}izales}, {Ros}, {Roshanineshat}, {Rottmann}, {Roy}, {Ruiz},
  {Ruszczyk}, {Rygl}, {S{\'a}nchez}, {S{\'a}nchez-Arg{\"u}elles},
  {S{\'a}nchez-Portal}, {Sasada}, {Satapathy}, {Savolainen}, {Schloerb},
  {Schonfeld}, {Schuster}, {Shao}, {Shen}, {Small}, {Sohn}, {SooHoo},
  {Sosapanta Salas}, {Souccar}, {Sun}, {Tazaki}, {Tetarenko}, {Tiede},
  {Tilanus}, {Titus}, {Torne}, {Toscano}, {Traianou}, {Trent}, {Trippe},
  {Turk}, {van Bemmel}, {van Langevelde}, {van Rossum}, {Vos}, {Wagner},
  {Ward-Thompson}, {Wardle}, {Washington}, {Weintroub}, {Wharton}, {Wielgus},
  {Wiik}, {Witzel}, {Wondrak}, {Wong}, {Wu}, {Yadlapalli}, {Yamaguchi},
  {Yfantis}, {Yoon}, {Young}, {Young}, {Younsi}, {Yu}, {Yuan}, {Yuan},
  {Zensus}, {Zhang}, {Zhao}, \& {Zhao}}]{M87_23}
{Event Horizon Telescope Collaboration}, {Akiyama}, K., {Alberdi}, A., {et~al.}
  2023, \apjl, 957, L20

\bibitem[{{Event Horizon Telescope Collaboration} {et~al.}(2019){Event Horizon
  Telescope Collaboration}, {Akiyama}, {Alberdi}, {Alef}, {Asada}, {Azulay},
  {Baczko}, {Ball}, {Balokovi{\'c}}, {Barrett}, {et~al.}}]{EHT_M87_PaperV}
{Event Horizon Telescope Collaboration}, {Akiyama}, K., {Alberdi}, A., {et~al.}
  2019, Astrophys. J. Lett., 875, L5

\bibitem[{{Event Horizon Telescope Collaboration} {et~al.}(2021){Event Horizon
  Telescope Collaboration}, {Akiyama}, {Algaba}, {Alberdi}, {Alef}, {Anantua},
  {Asada}, {Azulay}, {Baczko}, {Ball}, {Balokovi{\'c}}, {Barrett}, {Benson},
  {Bintley}, {Blackburn}, {Blundell}, {Boland}, {Bouman}, {Bower}, {Boyce},
  {Bremer}, {Brinkerink}, {Brissenden}, {Britzen}, {Broderick}, {Broguiere},
  {Bronzwaer}, {Byun}, {Carlstrom}, {Chael}, {Chan}, {Chatterjee},
  {Chatterjee}, {Chen}, {Chen}, {Chesler}, {Cho}, {Christian}, {Conway},
  {Cordes}, {Crawford}, {Crew}, {Cruz-Osorio}, {Cui}, {Davelaar}, {De
  Laurentis}, {Deane}, {Dempsey}, {Desvignes}, {Dexter}, {Doeleman}, {Eatough},
  {Falcke}, {Farah}, {Fish}, {Fomalont}, {Ford}, {Fraga-Encinas}, {Friberg},
  {Fromm}, {Fuentes}, {Galison}, {Gammie}, {Garc{\'\i}a}, {Gelles}, {Gentaz},
  {Georgiev}, {Goddi}, {Gold}, {G{\'o}mez}, {G{\'o}mez-Ruiz}, {Gu}, {Gurwell},
  {Hada}, {Haggard}, {Hecht}, {Hesper}, {Himwich}, {Ho}, {Ho}, {Honma},
  {Huang}, {Huang}, {Hughes}, {Ikeda}, {Inoue}, {Issaoun}, {James}, {Jannuzi},
  {Janssen}, {Jeter}, {Jiang}, {Jimenez-Rosales}, {Johnson}, {Jorstad}, {Jung},
  {Karami}, {Karuppusamy}, {Kawashima}, {Keating}, {Kettenis}, {Kim}, {Kim},
  {Kim}, {Kim}, {Kino}, {Koay}, {Kofuji}, {Koch}, {Koyama}, {Kramer}, {Kramer},
  {Krichbaum}, {Kuo}, {Lauer}, {Lee}, {Levis}, {Li}, {Li}, {Lindqvist}, {Lico},
  {Lindahl}, {Liu}, {Liu}, {Liuzzo}, {Lo}, {Lobanov}, {Loinard}, {Lonsdale},
  {Lu}, {MacDonald}, {Mao}, {Marchili}, {Markoff}, {Marrone}, {Marscher},
  {Mart{\'\i}-Vidal}, {Matsushita}, {Matthews}, {Medeiros}, {Menten}, {Mizuno},
  {Mizuno}, {Moran}, {Moriyama}, {Moscibrodzka}, {M{\"u}ller}, {Musoke}, {Mus
  Mej{\'\i}as}, {Michalik}, {Nadolski}, {Nagai}, {Nagar}, {Nakamura},
  {Narayan}, {Narayanan}, {Natarajan}, {Nathanail}, {Neilsen}, {Neri}, {Ni},
  {Noutsos}, {Nowak}, {Okino}, {Olivares}, {Ortiz-Le{\'o}n}, {Oyama},
  {{\"O}zel}, {Palumbo}, {Park}, {Patel}, {Pen}, {Pesce}, {Pi{\'e}tu},
  {Plambeck}, {PopStefanija}, {Porth}, {P{\"o}tzl}, {Prather},
  {Preciado-L{\'o}pez}, {Psaltis}, {Pu}, {Ramakrishnan}, {Rao}, {Rawlings},
  {Raymond}, {Rezzolla}, {Ricarte}, {Ripperda}, {Roelofs}, {Rogers}, {Ros},
  {Rose}, {Roshanineshat}, {Rottmann}, {Roy}, {Ruszczyk}, {Rygl},
  {S{\'a}nchez}, {S{\'a}nchez-Arguelles}, {Sasada}, {Savolainen}, {Schloerb},
  {Schuster}, {Shao}, {Shen}, {Small}, {Sohn}, {SooHoo}, {Sun}, {Tazaki},
  {Tetarenko}, {Tiede}, {Tilanus}, {Titus}, {Toma}, {Torne}, {Trent},
  {Traianou}, {Trippe}, {van Bemmel}, {van Langevelde}, {van Rossum}, {Wagner},
  {Ward-Thompson}, {Wardle}, {Weintroub}, {Wex}, {Wharton}, {Wielgus}, {Wong},
  {Wu}, {Yoon}, {Young}, {Young}, {Younsi}, {Yuan}, {Yuan}, {Zensus}, {Zhao},
  \& {Zhao}}]{2021ApJ...910L..13E}
{Event Horizon Telescope Collaboration}, {Akiyama}, K., {Algaba}, J.~C.,
  {et~al.} 2021, \apjl, 910, L13

\bibitem[{{Fishbone} \& {Moncrief}(1976)}]{1976ApJ...207..962F}
{Fishbone}, L.~G. \& {Moncrief}, V. 1976, \apj, 207, 962

\bibitem[{{Galishnikova} {et~al.}(2023){Galishnikova}, {Philippov}, {Quataert},
  {Bacchini}, {Parfrey}, \& {Ripperda}}]{2023PhRvL.130k5201G}
{Galishnikova}, A., {Philippov}, A., {Quataert}, E., {et~al.} 2023, \prl, 130,
  115201

\bibitem[{{Hakobyan} {et~al.}(2023){Hakobyan}, {Ripperda}, \&
  {Philippov}}]{2023ApJ...943L..29H}
{Hakobyan}, H., {Ripperda}, B., \& {Philippov}, A.~A. 2023, \apjl, 943, L29

\bibitem[{{Igumenshchev}(2008)}]{2008ApJ...677..317I}
{Igumenshchev}, I.~V. 2008, \apj, 677, 317

\bibitem[{{Katsoulakos} \& {Rieger}(2018)}]{2018ApJ...852..112K}
{Katsoulakos}, G. \& {Rieger}, F.~M. 2018, \apj, 852, 112

\bibitem[{{Komissarov}(2004)}]{2004MNRAS.350..427K}
{Komissarov}, S.~S. 2004, \mnras, 350, 427

\bibitem[{{Kusunose} \& {Mineshige}(1996)}]{1996ApJ...468..330K}
{Kusunose}, M. \& {Mineshige}, S. 1996, \apj, 468, 330

\bibitem[{{Levinson} \& {Rieger}(2011)}]{2011ApJ...730..123L}
{Levinson}, A. \& {Rieger}, F. 2011, \apj, 730, 123

\bibitem[{{Lu} {et~al.}(2023){Lu}, {Asada}, {Krichbaum}, {Park}, {Tazaki},
  {Pu}, {Nakamura}, {Lobanov}, {Hada}, {Akiyama}, {Kim}, {Marti-Vidal},
  {G{\'o}mez}, {Kawashima}, {Yuan}, {Ros}, {Alef}, {Britzen}, {Bremer},
  {Broderick}, {Doi}, {Giovannini}, {Giroletti}, {Ho}, {Honma}, {Hughes},
  {Inoue}, {Jiang}, {Kino}, {Koyama}, {Lindqvist}, {Liu}, {Marscher},
  {Matsushita}, {Nagai}, {Rottmann}, {Savolainen}, {Schuster}, {Shen}, {de
  Vicente}, {Walker}, {Yang}, {Zensus}, {Algaba}, {Allardi}, {Bach},
  {Berthold}, {Bintley}, {Byun}, {Casadio}, {Chang}, {Chang}, {Chang}, {Chen},
  {Chen}, {Chilson}, {Chuter}, {Conway}, {Crew}, {Dempsey}, {Dornbusch},
  {Faber}, {Friberg}, {Garc{\'\i}a}, {Garrido}, {Han}, {Han}, {Hasegawa},
  {Herrero-Illana}, {Huang}, {Huang}, {Impellizzeri}, {Jiang}, {Jinchi},
  {Jung}, {Kallunki}, {Kirves}, {Kimura}, {Koay}, {Koch}, {Kramer}, {Kraus},
  {Kubo}, {Kuo}, {Li}, {Lin}, {Liu}, {Liu}, {Lo}, {Lu}, {MacDonald},
  {Martin-Cocher}, {Messias}, {Meyer-Zhao}, {Minter}, {Nair}, {Nishioka},
  {Norton}, {Nystrom}, {Ogawa}, {Oshiro}, {Patel}, {Pen}, {Pidopryhora},
  {Pradel}, {Raffin}, {Rao}, {Ruiz}, {Sanchez}, {Shaw}, {Snow}, {Sridharan},
  {Srinivasan}, {Tercero}, {Torne}, {Traianou}, {Wagner}, {Walther}, {Wei},
  {Yang}, \& {Yu}}]{Lu23}
{Lu}, R.-S., {Asada}, K., {Krichbaum}, T.~P., {et~al.} 2023, \nat, 616, 686

\bibitem[{{McKinney}(2006)}]{2006MNRAS.368.1561M}
{McKinney}, J.~C. 2006, \mnras, 368, 1561

\bibitem[{{Mo{\'s}cibrodzka} {et~al.}(2011){Mo{\'s}cibrodzka}, {Gammie},
  {Dolence}, \& {Shiokawa}}]{2011ApJ...735....9M}
{Mo{\'s}cibrodzka}, M., {Gammie}, C.~F., {Dolence}, J.~C., \& {Shiokawa}, H.
  2011, \apj, 735, 9

\bibitem[{{Muslimov} \& {Harding}(1997)}]{1997ApJ...485..735M}
{Muslimov}, A. \& {Harding}, A.~K. 1997, \apj, 485, 735

\bibitem[{{Narayan} {et~al.}(2003){Narayan}, {Igumenshchev}, \&
  {Abramowicz}}]{2003PASJ...55L..69N}
{Narayan}, R., {Igumenshchev}, I.~V., \& {Abramowicz}, M.~A. 2003, \pasj, 55,
  L69

\bibitem[{{Ng} {et~al.}(2024){Ng}, {Jiang}, {Musolino}, {Ecker}, {Tootle}, \&
  {Rezzolla}}]{2024PhRvD.109f4061N}
{Ng}, H. H.-Y., {Jiang}, J.-L., {Musolino}, C., {et~al.} 2024, \prd, 109,
  064061

\bibitem[{{Olivares} {et~al.}(2019){Olivares}, {Porth}, {Davelaar}, {Most},
  {Fromm}, {Mizuno}, {Younsi}, \& {Rezzolla}}]{2019A&A...629A..61O}
{Olivares}, H., {Porth}, O., {Davelaar}, J., {et~al.} 2019, \aap, 629, A61

\bibitem[{{Papadopoulos} \& {Contopoulos}(2019)}]{Papadopoulos2019}
{Papadopoulos}, D.~B. \& {Contopoulos}, I. 2019, Mon. Not. R. Astron. Soc.,
  483, 2325

\bibitem[{{Paraschos} {et~al.}(2024){Paraschos}, {Debbrecht}, {Kramer},
  {Traianou}, {Liodakis}, {Krichbaum}, {Kim}, {Janssen}, {Nair}, {Savolainen},
  {Ros}, {Bach}, {Hodgson}, {Lisakov}, {MacDonald}, \& {Zensus}}]{P24b}
{Paraschos}, G.~F., {Debbrecht}, L.~C., {Kramer}, J.~A., {et~al.} 2024, \aap,
  686, L5

\bibitem[{{Paraschos} {et~al.}(2023){Paraschos}, {Mpisketzis}, {Kim}, {Witzel},
  {Krichbaum}, {Zensus}, {Gurwell}, {L{\"a}hteenm{\"a}ki}, {Tornikoski},
  {Kiehlmann}, \& {Readhead}}]{P23}
{Paraschos}, G.~F., {Mpisketzis}, V., {Kim}, J.~Y., {et~al.} 2023, \aap, 669,
  A32

\bibitem[{{Petropoulou} {et~al.}(2019){Petropoulou}, {Yuan}, {Chen}, \&
  {Mastichiadis}}]{2019ApJ...883...66P}
{Petropoulou}, M., {Yuan}, Y., {Chen}, A.~Y., \& {Mastichiadis}, A. 2019, \apj,
  883, 66

\bibitem[{{Porth} {et~al.}(2021){Porth}, {Mizuno}, {Younsi}, \&
  {Fromm}}]{2021MNRAS.502.2023P}
{Porth}, O., {Mizuno}, Y., {Younsi}, Z., \& {Fromm}, C.~M. 2021, \mnras, 502,
  2023

\bibitem[{{Porth} {et~al.}(2017){Porth}, {Olivares}, {Mizuno}, {Younsi},
  {Rezzolla}, {Moscibrodzka}, {Falcke}, \& {Kramer}}]{2017ComAC...4....1P}
{Porth}, O., {Olivares}, H., {Mizuno}, Y., {et~al.} 2017, Computational
  Astrophysics and Cosmology, 4, 1

\bibitem[{{Ptitsyna} \& {Neronov}(2016)}]{2016A&A...593A...8P}
{Ptitsyna}, K. \& {Neronov}, A. 2016, \aap, 593, A8

\bibitem[{{Ripperda} {et~al.}(2022){Ripperda}, {Liska}, {Chatterjee}, {Musoke},
  {Philippov}, {Markoff}, {Tchekhovskoy}, \& {Younsi}}]{2022ApJ...924L..32R}
{Ripperda}, B., {Liska}, M., {Chatterjee}, K., {et~al.} 2022, \apjl, 924, L32

\bibitem[{{Stathopoulos} {et~al.}(2024){Stathopoulos}, {Petropoulou}, {Sironi},
  \& {Giannios}}]{2024arXiv240601211S}
{Stathopoulos}, S.~I., {Petropoulou}, M., {Sironi}, L., \& {Giannios}, D. 2024,
  arXiv e-prints, arXiv:2406.01211

\bibitem[{{Tchekhovskoy} {et~al.}(2011){Tchekhovskoy}, {Narayan}, \&
  {McKinney}}]{2011MNRAS.418L..79T}
{Tchekhovskoy}, A., {Narayan}, R., \& {McKinney}, J.~C. 2011, \mnras, 418, L79

\bibitem[{{Vincent} \& {Lebohec}(2010)}]{2010MNRAS.409.1183V}
{Vincent}, S. \& {Lebohec}, S. 2010, \mnras, 409, 1183

\end{thebibliography}

\label{lastpage}
\end{document}